\documentclass[twocolumn]{revtex4}
\usepackage{amsmath, amsfonts, amssymb, graphicx, color, soul}

\graphicspath{{./figures}}

\newcommand{\beq}{\begin{equation}}
\newcommand{\eeq}{\end{equation}}
\newcommand{\beqs}{\begin{equation*}}
\newcommand{\eeqs}{\end{equation*}}
\newcommand{\beqn}{\begin{eqnarray}}
\newcommand{\eeqn}{\end{eqnarray}}
\newcommand{\beqns}{\begin{eqnarray*}}
\newcommand{\eeqns}{\end{eqnarray*}}

\newcommand{\cct}{\mathbf{G}\mathbf{G}^\intercal}
\newcommand{\cz}{\mathbf{C}(0)}
\newcommand{\ctau}{\mathbf{C}(\tau)}
\newcommand{\coh}{\text{Coh}}

\newcommand{\RV}{}

\begin{document}

\title{Inferring resource competition in microbial communities from time series}

\author{Xiaowen Chen$^{1}$, Kyle Crocker$^{2,3}$, Seppe Kuehn$^{2,3,4,5,*}$, Aleksandra M. Walczak$^{1,4,*}$ and Thierry Mora$^{1,4,*}$}

\affiliation{$^{1}$ Laboratoire de Physique de l'\'Ecole normale sup\'erieure, ENS, Universit\'e PSL, CNRS, Sorbonne Universit\'e, Universit\'e Paris Cit\'e, F-75005 Paris, France}
\affiliation{$^{2}$ Department of Ecology and Evolution, The University of Chicago, Chicago,  USA}
\affiliation{$^{3}$ Center for the Physics of Evolving Systems, The University of Chicago, Chicago, USA}
\affiliation{$^{4}$ Center for Living Systems, The University of Chicago, Chicago, USA}
\affiliation{$^{5}$ National Institute for Theory and Mathematics in Biology, The University of Chicago and Northwestern University, Chicago, USA}
\affiliation{$^*$  Corresponding authors. The authors contributed equally to this work.}

\begin{abstract}
The competition for resources is a defining feature of microbial communities. In many contexts, from soils to host-associated communities, highly diverse microbes are organized into metabolic groups or guilds with similar resource preferences. The resource preferences of individual taxa that give rise to these guilds are critical for understanding fluxes of resources through the community and the structure of diversity in the system. However, inferring the metabolic capabilities of individual taxa, and their competition with other taxa, within a community is challenging and unresolved. Here we address this gap in knowledge by leveraging dynamic measurements of abundances in communities. We show that simple correlations are often misleading in predicting resource competition. We show that spectral methods such as the cross-power spectral density (CPSD) and coherence that account for time-delayed effects are superior metrics for inferring the structure of resource competition in communities. We first demonstrate this fact on synthetic data generated from consumer-resource models with time-dependent resource availability, where taxa are organized into groups or guilds with similar resource preferences. By applying spectral methods to oceanic plankton time-series data, we demonstrate that these methods detect interaction structures among species with similar genomic sequences. Our results indicate that analyzing temporal data across multiple timescales can reveal the underlying structure of resource competition within communities.
\end{abstract}

\maketitle

\section{Introduction}

The collective biological activity of ecosystems is defined by fluxes of resources. From plant~\cite{wilson_plant_1993} to microbial~\cite{foster_competition_2012} communities, the availability of resources such as sunlight or reduced carbon and nitrogen enables the generation of energy and the production of biomass. Competition for resources therefore drives ecological interactions between members of the system. As a result, resource-mediated interactions such as competition have long been recognized as central structuring properties of ecosystems across scales.

Perhaps nowhere are the structuring forces of resources more clear than in microbial consortia. Communities of microbes utilize resources in almost every niche on the planet, from carbon remineralization in the photic zone of the ocean~\cite{datta_microbial_2016}, to fiber utilization in the rumen of ungulates~\cite{shabat_specific_2016}, and the collective transformation of oxidized nitrogen compounds in the soil~\cite{tiedje_denitrification_1983,gowda_genomic_2022}. In each of these cases, complex communities of hundreds or thousands of species compete for, and utilize, resources to generate energy and biomass. 

In these contexts, the metabolic traits of microbes utilizing each resource become a key property of interest because they dictate which species compete for which resources. In part due to the conserved structure of the biochemical pathways that enable the utilization of resources, microbial communities are frequently observed to be comprised of guilds, or groups of species, that utilize a similar set of resources~\cite{louca_decoupling_2016,nelson_global_2016}. These guilds, often comprised of tens or more taxa, compete for diverse pools of resources. For example, marine communities utilize complex mixtures of carbon sources~\cite{kharbush_particulate_2020} and many members of the human gut microbiome utilize mixtures of carbon sources~\cite{tramontano_nutritional_2018}. Thus the mapping of species to the resources they utilize typically has a block-like structure, with groups of species utilizing groups of resources~\cite{louca_function_2018}.

A second key feature of resource utilization in microbial ecosystems is temporal fluctuations. Temporal variability in resource availability is the norm across many if not most habitats. For example, diurnal fluctuations drive the availability of carbon in photosynthetically driven ecosystems~\cite{staley_diurnal_2017}, and feeding temporally structures the availability of resources in host-associated microbial communities~\cite{salari_diurnal_2024}. Similarly, in soils, transient dynamics in moisture drive large fluctuations in nutrient availability~\cite{birch_effect_1958}. 

Given the importance of resources, the central role of microbial traits and temporal fluctuations in mediating the dynamic utilization of resources by communities, a central challenge in the field is understanding resource-mediated interactions in communities. This challenge is highlighted by the fact that in the context of a complex community, we cannot easily access the resources each and every species is utilizing. In most cases, isolating individual strains or taxa and assaying their resource preferences is infeasible. While genome-scale models~\cite{klitgord_environments_2010} show potential for predicting resource utilization from sequencing data (shotgun metagenomics) these methods remain highly error-prone for genomes of non-model organisms~\cite{li_statistical_2023}. In light of these considerations, it is important to develop empirically grounded methods to infer the resource-mediated interactions in a complex consortium.

Here, we ask what metrics on the observed time series of species abundances can give us insight into the structure of resource-mediated interactions in the community. In particular, if there are metabolic guilds in the community, how can we reliably infer them from time-series measurements?  
We approach this problem in the context of consumer-resource models~\cite{macarthur_species_1970}. We focused on consumer-resource formalisms because recent work has shown these models to be powerful quantitative formalisms for understanding abundance and resource dynamics in real communities~\cite{gowda_genomic_2022,ho_resource_2024}, suggesting that this formalism has some predictive power in the microbial context.

While a host of powerful methods have been developed to infer interactions in ecological communities many of these methods rely on statistical analysis of correlations in species abundances and make simplifying assumptions such as sparsity~\cite{spieceasi,friedman_inferring_2012}. Leveraging the widespread availability of time series data in microbial communities~\cite{robicheau_highly-resolved_2022,marine_data,goyal_interactions_2022}, we present temporal pairwise measurements as novel predictors of resource-mediated species-species interactions and apply them to existing datasets.

\section{Results}

\subsection{Environment-mediated consumer-resource model}

\begin{figure*}
\includegraphics[scale=1]{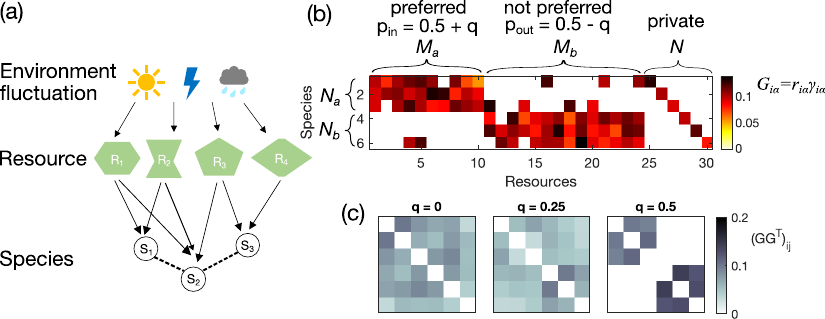}
\caption{{\bf Environment-mediated ecological interactions between resources and consumers.} (a)  Cartoon of interactions among environment-mediated resources and consumer species. (b) An example of the gain matrix for consumer-resource interaction $\mathbf{G}$ with elements $G_{i\alpha} = r_{i\alpha}\gamma_{i\alpha}$ (colorbar).  The probability of non-zero consumer-resource interactions depends on whether the resource is preferred by the species or not, with an adjustable parameter $q$ that encodes guild-structure bias ($q = 0.4$ for this example). 
All resources also have private resources to prevent extinction.  (c) Effective interactions among the species are captured by  the effective resource-utilization overlap matrix, $\cct$ (colorbar), shown here for 3 different values of the guild-structure bias $q$.}
\label{fig:crm_schematics}
\end{figure*}

In real experiments, the underlying consumer-resource interactions and resource dynamics are exceedingly challenging to measure. To develop a method that can be applied to analyze experimental time traces, we first introduce a model with known ground truth to develop an appropriate metric, which can reveal the effective resource overlap between species, using only the observed abundances of consumer species.

We consider an ecological system with $N$ species with abundances $x_i$ ($i = 1, \dots N$) and $M$ resources with concentrations $R_\alpha$ ($\alpha = 1, \dots M$). Interactions are given by an externally supplied resource model. The dynamical effect of environmental fluctuations is explicitly modeled by the time-dependence of the external supply, $K_\alpha(t)$,  that mediates  resource availability  (Fig.~\ref{fig:crm_schematics}(a))~\cite{macarthur_species_1970,cui_effect_2020,marsland_community_2020,advani_statistical_2018}. 
Mathematically,
\begin{eqnarray}
\dot{x}_i & = & \left( \sum_{\alpha=1}^M r_{i\alpha} \gamma_{i\alpha}  R_\alpha(t)  - d_x\right) x_i(t), \label{crm1}\\ 
\dot{R}_\alpha & = & K_\alpha(t) -\left( d_R  + \sum_{i=1}^N r_{i\alpha} x_i(t) \right) R_\alpha(t),\label{crm2}
\end{eqnarray}
where $r_{i\alpha}$ is the uptake rate of resource $\alpha$ for species $i$ (and simultaneously the depletion rate of resource $\alpha$ given species $i$), $\gamma_{i \alpha}$ is the yield of species $i$ given resource $\alpha$, i.e. how much species $i$ grows by consuming resource $\alpha$,  
and $d_x$ and $d_R$ are the death rate of the species and the resources. 
For simplicity, $d_x$ is set to be equal for all species and $d_R$ across all resources. 
We call this model the environment-mediated consumer-resource model (ECRM). 

Because resources impact species abundances through exponential growth, it will be useful to log-transform species abundances before analysis, denoting
\begin{equation}
\tilde{s}_i(t) \equiv \log x_i(t),
\end{equation}
which we further normalize as the \textit{z}-score,
\begin{equation}
s_i(t) = \frac{\tilde{s}_i(t) - \langle \tilde{s}_i \rangle }{\sigma_s}.
\end{equation}
Intuitively, the $z$-score describe the deviation from the mean in terms of expected standard deviation.

To model the impact of environment, we assume that its fluctuations directly influence the resource abundance, which then by means of the consumer-resource relation is transmitted through the ecological system.
We consider two types of environmental fluctuations, $K_\alpha(t)$.

First, sinusoidal drives are given by
\begin{equation}\label{sindrive}
K_\alpha(t) = K_\alpha^0 +  A_\alpha \sin(\omega_\alpha t - \phi_\alpha),
\end{equation}
where $K_\alpha^0$ is the base supply for resource $\alpha$, $A_\alpha$ the amplitude of the sinusoidal drive, $\omega_\alpha$ the frequency, and $\phi_\alpha$ the phase. This form models the periodic drives omnipresent in nature, for example, diurnal cycles in aquatic systems, or regular food intakes in host-associated microbiomes.

Second, Ornstein-Uhlenbeck (OU) drives evolve according to
\begin{equation}\label{oudrive1}
K_\alpha(t) = K_\alpha^0 + Q_\alpha(t), 
\end{equation}
with $K_\alpha^0$ the base supply for resource $\alpha$ as in the sinusoidal case and $Q_\alpha$ a stochastic OU process defined by
\begin{equation}\label{oudrive2}
\frac{dQ_\alpha}{dt} = - \omega_\alpha Q_\alpha + A_\alpha \sqrt{\omega_\alpha} \eta_\alpha(t). 
\end{equation}
Here, $\omega_\alpha$ and $ A_\alpha$ are defined such that both the power and the timescales in the sinusoidal and the OU processes match, and the white noise $\lbrace\eta_\alpha(t)\rbrace$ satisfies
\begin{equation}
\langle \eta_\alpha(t) \rangle = 0, \hspace{0.5cm}
\langle \eta_\alpha(t) \eta_\beta(t') \rangle = \delta_{\alpha \beta} \delta(t-t').
\end{equation}
Compared to sinusoidal drives, OU drives are stochastic, which are useful in modeling environmental fluctuations with unknown strength and regularity such as redox fluctuations in soils~\cite{liptzin_temporal_2011}.

The gain matrix for the consumer-resource interaction $\mathbf{G}$, where each entry is $G_{i\alpha} = r_{i \alpha} \gamma_{i \alpha}$, describes the intake of resource $\alpha$ by species $i$ (Fig.~\ref{fig:crm_schematics}(b)). 
Summing over the resources we obtain the resource-utilization overlap matrix $\cct$ (Fig.~\ref{fig:crm_schematics}(c)). Large off-diagonal entries of this matrix ($\cct_{i,j}$) indicate that species $i$ and $j$ have high overlap in the resources they utilize. Species with high resource-overlap are expected to compete, and those without are expected not to compete. One can consider the resource-utilization matrix as a proxy for effective interactions between species, which is informative about the community structure. 

Specifically, we are interested in detecting guilds among the interacting species. 
To introduce a guild structure in resource overlap, we write the uptake rate as $r_{i\alpha} = r g_{i\alpha}$, where $g_{i\alpha}$ is an adjacency matrix that takes the value of 0 if species $i$ does not uptake resource $\alpha$, and 1 if it does consume this resource. 
In order to generate ensembles of random interaction strengths such that the resulting $\cct$ matrix has a block (guild) structure, we construct a bipartite graph between $N$ species and $M$ resources. We then separate both the species and the resources into $k$ groups, with each group of species preferring to consume resources from one group of resources. The elements of the adjacency matrix that specifies these preferences, $g_{i\alpha}$, are drawn from a Bernoulli distribution parametrized by a probability $p_\text{in}$. For the other $k-1$ groups of non-preferred resources, the elements of the adjacency matrix $g_{i\alpha}$ are drawn from a Bernoulli distribution parametrized by a  probability $p_\text{out}<p_{\rm in}$. 
In addition, each species is assigned its private resource to prevent extinction (Fig.~\ref{fig:crm_schematics}(b)). The yield $\gamma_{i\alpha}$ is drawn from a Gaussian distribution with a positive mean and small variance to introduce randomness. All yields are non-negative (see \textbf{Methods} for specific parameter values).

When there are only two guilds $a$ and $b$, we can reduce the number of parameters by defining a guild-structure bias $q$ and set $p_\text{in} = 0.5 + q$ and $p_\text{out} = 0.5 - q$. For the first set of $N_{a}$ species we set the probability that they consume the $M_a$ resources to $p_\text{in}= 0.5 + q$, while their probability to consume the remaining $M_b$ resources is $p_\text{out}= 0.5 - q$. The $N_b$ species from the second group have opposite preferences: with probability $p_\text{out} = 0.5 - q$ they consume the first $M_a$ resources, and probability $p_\text{in} = 0.5 + q$ the remaining $M_b$ resources.  Here, $q$ acts as a tuning variable for block structures. If $q = 0.5$, there are zero interactions between species that belong to different guilds. If $q = 0$, there is no bias of any species towards any  resource (see Fig.~\ref{fig:crm_schematics}(b,c)). 

\begin{figure*}
\includegraphics[width=0.8\textwidth]{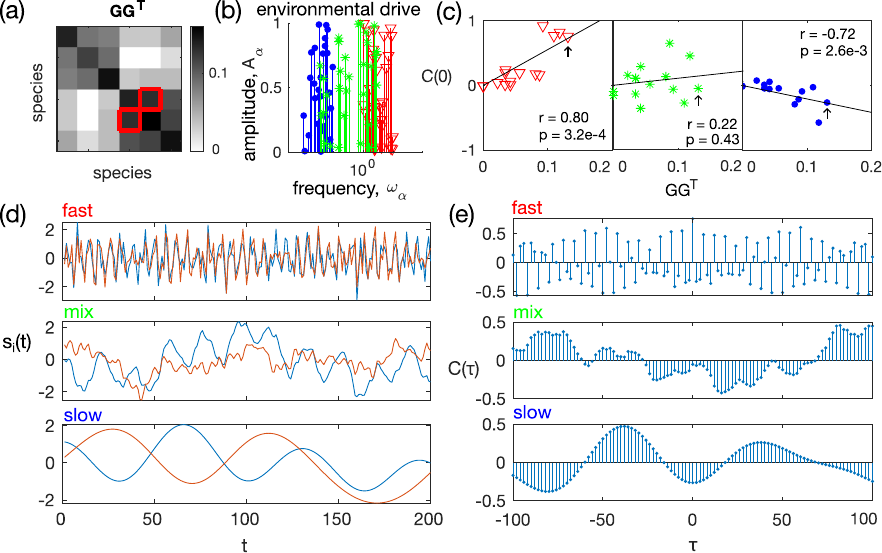}
\caption{{\bf The simulated guild based ecosystem.} An example environment-mediated consumer-resource model (number of species $N = 6$, number of resources $M = 30$, and guild-structure bias $q = 0.4$) demonstrating that the relation between the equal-time species abundance correlation $\cz$ and the underlying resource utilization overlap matrix $\cct$ depend on the timescale of the environmental drive. (a) Resource utilization overlap matrix $\cct$. The red boxes indicate an example pair of interacting species. (b) Environmental drives are chosen to be sinusoidal, with fast (\textit{red}, $\omega_\alpha\in \left[1, 10\right]$), slow (\textit{blue}, $\omega_\alpha\in \left[0.01, 0.1\right]$), and mixed timescales (\textit{green}, $\omega_\alpha\in \left[0.03, 3\right]$). (c) The correlation between the elements of equal-time correlation function $\cz$ and the elements of the resource-utilization matrix, $\cct$, is positive when the environmental drive is fast (\textit{red}); negative when the drive is slow (\textit{blue}), and shows no correlation when the drive is composed of a mixture of slow and fast timescales (\textit{green}). In each panel, $r$ is Spearman's rank correlation coefficient, $p$ is the p-value for the null hypothesis that there is no correlation between $\cz$ and $\cct$. Black arrows indicate the example pair of species in the red boxes of (\textit{a}). (d) Example segments of time series of the log-transformed species abundances for the red-boxed pair of (\textit{a}), under the three types of environmental drive given in panel (\textit{b}). (e) Cross-correlation function between the species abundances of the example red-boxed pair of (\textit{a}). Equal-time correlation is the value of $\ctau$ at $\tau = 0$.
}\label{fig:corr_vs_cct}
\end{figure*}

The consumer-resource model \eqref{crm1}-\eqref{crm2}, environmental drive \eqref{sindrive}-\eqref{oudrive2}, and resource guild structure $r_{i\alpha}$ among consumers, completely define the model.
To illustrate its behavior and gain intuition, we first simulated a small network with $N = 6$ species and $M = 30$ resources, with a guild-structure bias of $q = 0.4$. The resource-utilization matrix $\cct$ is given by Fig.~\ref{fig:corr_vs_cct}(a). The specific set of parameters is given in \textbf{Methods}.
The first $N_a = 3$ species form one guild, while the remaining $N_b=3$ species form the other. To study the impact of environmental drive on species abundances, we simulate the ECRM under sinusoidal and OU environmental drives each sampled at three different ranges of timescales -- slow, fast, and with a mixture of timescales (Fig.~\ref{fig:corr_vs_cct}(b)). Example traces of pairs of species within the same guild show complex coupled dynamics depending on the environmental timescales  (Fig.~\ref{fig:corr_vs_cct}(d)).

\subsection{Pairwise measures of species couplings}

To explore how the coupled dynamics of species abundances can be informative about their resource overlap $\cct$, we consider observables of increasing complexity that measure the deviation from independence between the time traces of two species. 

\subsubsection{Equal-time correlation $\cz$} 

We first consider the equal-time correlation coefficient, defined  as
\begin{equation}
C_{ij}(0) = \lim_{T \rightarrow \infty}
\frac{1}{T}\int_{0}^{T} s_i(t) s_j(t) dt,
\end{equation}
which for finite data time-series size sampled at a finite sampling rate is
\begin{equation}
    C_{ij}(0) \approx \frac{1}{T}\sum_{t = 1}^T s_i(t) s_j(t),
\end{equation}
where $t=1,\ldots,T$ now denotes the time frame.
As equal-time correlations describes how much the concentrations of pairs of species fluctuate together, it is a measure commonly used in detecting species-species interactions. However, it is known that equal-time correlations cannot accurately reproduce the interactions~\cite{friedman_inferring_2012}. 

Specifically, as shown in Figure~\ref{fig:corr_vs_cct}(c) in our example of 6 species distributed into 2 guilds, when the environmental drive is fast, $C_{ij}(0) \propto \left(\cct\right)_{ij}$, while when the environmental drive is slow, $C_{ij}(0) \propto - \left(\cct\right)_{ij}$. Intuitively this happens because fast resource dynamics drive coherent responses of members within a guild, but slow resource dynamics lead to competition within guilds (the underlying mechanism is first explored and explained in detail in~\cite{crocker_microbial_2025}). 
This problem is more pronounced when  environmental fluctuations are driven by a mixture of timescales (middle panel in Fig.~\ref{fig:corr_vs_cct}(c)). There is no correlation between the equal-time correlation $\cz$ and the resource utilization overlap  $\cct$ ($p$-value = 0.4, Spearman correlation test).  In this example, the pair of species with the strongest resource utilization overlap (indicated by red boxes in Fig.~\ref{fig:corr_vs_cct}(a)) has a equal-time correlation close to 0, as indicated by the black arrow in Fig.~\ref{fig:corr_vs_cct}(c).

Without knowing \textit{a priori} the timescale of the environmental fluctuation, the equal-time correlation $\cz$ does not return a good enough estimator of the resource utilization overlap. 

\begin{figure*}
\includegraphics[width=1\textwidth]{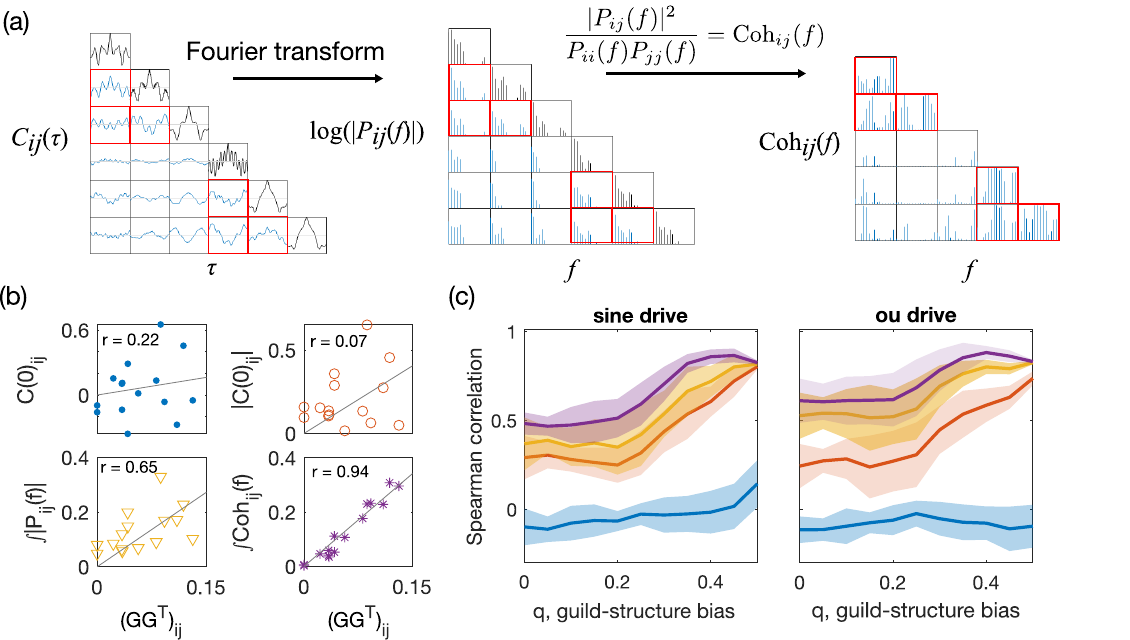}
\caption{\textbf{Dynamical pairwise observables reveal more about the resource utilization overlap structure than equal-time correlations.} (a) A schematic representation for calculating dynamical observables from the cross-correlation function. The pairwise cross-correlation function (left) is Fourier-transformed into the cross power spectral density (CPSD) (middle), and squared and normalized to obtained the squared coherence (right). All plots show results for the ECRM example in Fig.~\ref{fig:corr_vs_cct} under an environmental drive with mixed timescales. Pairs of species within the same guild are indicated by red boxes. The guild-structure bias is $q = 0.4$. 
(b) Scatter plots between pairwise observables for each pair of species and the elements of the effective resource-utilization matrix $\cct$ for same data as in panel (a). 
Pairwise observables include the equal time correlation $\cz$ (reproduced from Fig.~\ref{fig:corr_vs_cct}(b) \textit{middle panel}),  its absolute value $\lvert\cz\rvert$,  the total magnitude of CPSD ($\mathcal{P}_{ij} = \int df \lvert P_{ij}(f)\rvert $), and the total coherence ($\mathcal{TC}_{ij} = \int df \coh_{ij}(f)$).  $r$ is the Spearman's rank correlation coefficient. 
(c) Prediction performance of the pairwise measures of species abundance for resource utilization overlap, given by Spearman's correlation coefficient between the measure and the resource-utilization matrix value, as a function of the guild-structure bias $q$ (same color code as panel (\textit{b})). Shaded areas represent standard deviation across 10 realizations of the network, each with 10 random realizations of environmental drive given by a sinusoidal process with $\omega_\alpha$ sampled log-uniformly $ \in \left[0.03, 3\right]$ (\textit{left}), or by an OU process with $\omega_\alpha = 1$ (\textit{right}). 
The total duration of the simulated trajectory is $t_f = 20000$. The sampling timestep is $\Delta t_\text{sampling} = 1$.}
\label{fig:temp_vs_q}
\end{figure*}

\subsubsection{Cross-correlation with time-delay reveals species interaction structure}

Since equal-time correlations do not reveal the structure of effective species-species interactions, we took a closer look at time series data. Let us focus again on the pair of species with the strongest resource overlap (red box in Fig.~\ref{fig:corr_vs_cct}(a)). As shown by Fig.~\ref{fig:corr_vs_cct}(d), the time evolution of species  abundances depends heavily on the environmental drive. Under fast environmental drive, the two species abundances  track each other closely, while under slow drive, they exhibit a phase shift. When the environmental drive is mixed with fast and slow modes, the dynamics of each species also exhibit a mixture of modes, and the relation between the two species is less clear. 

Given a complicated time-series, we need to consider dynamical pairwise measures, such as the time-delayed cross-correlation function,
\begin{equation}
C_{ij}(\tau) = \lim_{T \rightarrow \infty}
\frac{1}{T}\int_{0}^{T} s_i(t) s_j(t+\tau) dt.
\end{equation}
As shown in Figure~\ref{fig:corr_vs_cct}(e), under all three types of environmental drive, the absolute value of the cross-correlation at certain time lags is large, which suggests that it carries information about the species--species interaction structure.  
However, if one only considers a single time delay, e.g. $\tau = 0$, the value can be very close to 0.

Comparing across different pairs of species, we observe that the cross-correlation function between species of the same guild varies with a much larger amplitude than between species from a different guild (Fig.~\ref{fig:temp_vs_q}(a) \textit{left panel}, pairs of distinct species belonging to the same guild are outlined in red). 
However, different pairs of interacting species show signatures of strong cross-correlations at different time delays $\tau$, implying that there is no \textit{a priori} choice of any single time delay that would reveal the interaction structure between all pairs of species. 

\subsubsection{Spectral method for time-series analysis: Cross power spectral density (CPSD) and coherence}

In order to focus on the interplay of timescales, we re-formulated the dynamical pairwise measures in the frequency domain. Spectral analysis has been used to probe many aspects of ecological systems~\cite{platt1975spectral}. Here, we measure the cross power spectral density (CPSD) by taking the Fourier transform of the cross-correlation function,
\begin{equation}
P_{ij}(f) = \int_{-\infty}^\infty C_{ij}(\tau) e^{-i 2\pi f \tau} d\tau. 
\end{equation}
We divide its magnitude-squared by the power spectral density of each of the two time series, to define the (magnitude-squared) coherence, 
\begin{equation}
\coh_{ij}(f) = \frac{\lvert P_{ij}(f) \rvert^2}{P_{ii}(f) P_{jj}(f)}. 
\end{equation}
Coherence ranges between 0, when the two signals are unrelated, and 1, when they are perfectly correlated up to a phase shift. 
It is used in signal processing to detect relations between two signals~\cite{coherence_book_2023}.

Figure~\ref{fig:temp_vs_q}(a) shows a schematic of how to compute CPSD and coherence from the cross-correlation function for the example network in Fig.~\ref{fig:corr_vs_cct}(a) with mixed timescales of the environmental drive. 
Recalling that the underlying resource utlization overlap network has a block structure (the first three species belong to the same guild, and the last three species belong to another guild), it becomes clear from Fig.~\ref{fig:temp_vs_q}(a) that both the magnitude of the CPSD and the coherence are larger between intra-guild pairs of species than between inter-guild species. As coherence further reduces the impact of individual species (due to normalization), the difference between intra- and inter-guild pairs is larger in the coherence than in the  CPSD magnitude (Fig.~\ref{fig:corr_vs_cct}(a)).

Both CPSD and coherence are frequency dependent and contain information about the relation between the  species abundance \textit{z}-scores $s_i(t)$ and $s_j(t)$ at each frequency. 
To obtain an aggregate measure of the coupling between species abundances over many timescales, we consider the integral of the CPSD magnitude and coherence across all frequencies.
Specifically,
\begin{equation}
\mathcal{P} \equiv    \int_0^{f_\text{max}}  \lvert P_{ij}(f) \rvert df \approx \sum_{k = 0}^{k_{\rm max}} \lvert P_{ij}(f_k) \rvert \Delta f,
\end{equation}
where $ \Delta f$ is 
the frequency resolution we choose when performing the discrete Fourier transform. Note that if we integrated the CPSD without taking its magnitude, we would recover the equal-time correlation by Parseval's theorem.

Likewise, we define the total coherence as the integral of the magnitude squared coherence across all frequencies,  
\begin{equation}
\mathcal{TC}_{ij} \equiv \int_0^{f_\text{max}}  \coh_{ij}(f) df \approx \sum_{k = 0}^{k_{\rm max}} \coh_{ij}(f_k) \Delta f.
\end{equation}
The maximum frequency is given by the Nyquist frequency, $f_\text{max} = 1/2\Delta t_\text{sampling}$. The frequency resolution $\Delta f$ is chosen heuristically, such that the power spectral density estimate is relatively stable when $\Delta f$ is doubled or halved ($\Delta f = f_\text{max}/15$ for all the results presented in this manuscript).
Using the simulated ECRM data, we estimate the power spectral density of the species abundance time series using the MVGC Toolbox~\cite{mvgc_toolbox} and the MATLAB built-in function with Welch's method to reduce noise from using finite data. As shown in Fig.~\ref{fig:temp_vs_q}(b), compared to the equal-time correlation $\cz$ and its absolute value, both the total magnitude of CPSD $\mathcal{P}$ and the total coherence $\mathcal{TC}$ are much more informative of the underlying resource utilization overlap matrix $\cct$.
Total coherence is the best of the four predictors. This advantage of the total coherence persists for both ECRMs under sinusoidal and OU drives and over the range of guild-structure biases $q$ (Fig.~\ref{fig:temp_vs_q}(c)). 

Thus, the two proposed measures $\mathcal{P}$ and $\mathcal{TC}$ offer a proxy for the resource-overlap interactions between species.

\subsection{Binary classifier for guild-structure detection}

In many cases, one is interested in recovering the guild structure of the community from time series~\cite{shan_annotation-free_2023}. Identification of guild structure can enable a massive dimensional reduction in the dynamic description of the system and provide insight into key metabolic properties~\cite{lee_functional_2024}.  

Motivated by this, we asked whether total coherence and CPSD magnitude could be used to learn the large-scale guild structure of $\cct$.
We used the four pairwise measures, $\cz$, $\lvert \cz \rvert$, $\mathcal{P}$,  $\mathcal{TC}$, as scores to perform a binary classification between pairs of species belonging to the same guild and pairs belonging to different guilds. We applied this approach to our small $N=6$ simulated ECRM system with 2 guilds (Fig.~\ref{fig:corr_vs_cct}(a)). 
To obtain a consistent partition into guilds, we perform single linkage clustering initiated from all pairs of species whose pairwise measure is larger than a threshold within the range of the pairwise measure.
For example, if after thresholding species $i$ and $j$ are determined to belong to the same guild, as well as species $j$ and $k$, then all three species $i$, $j$, $k$ are grouped into to the same guild.

\begin{figure}
\includegraphics[width=1\columnwidth]{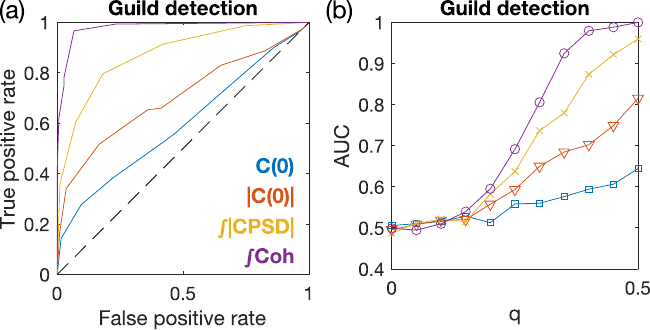}
\caption{\textbf{Guild detection of resource utilization overlap using the pairwise measures of species abundances with single linkage clustering. }
(a) Receiver Operator Curve (ROC) for successfully distinguishing same-guild from different-guild pairs of species using the 4 pairwise measures, from 50 random realizations of the simulated model in Fig.~\ref{fig:corr_vs_cct}. The guild-structure bias is set to $q = 0.4$.  (b) Area under the ROC (AUC) for guild detection. Same color code as (a). Environmental drives are given by OU processes with $\omega_\alpha = 1$.
Total duration is $t_f = 20000$. Sampling time step $\Delta t_\text{sampling} = 1$.}
\label{fig:roc_auc}
\end{figure}

To visualize the performance of the binary classifiers, we plot the Receiver Operator Curve (ROC) across different thresholds (Fig.~\ref{fig:roc_auc}(a)). The total coherence outperforms all other measures in correctly partitioning species into guilds. The Area Under the ROC (AUC) shows the advantage of coherence persists for all guild biases $q$ (Fig.~\ref{fig:roc_auc}(b)). For $q=0$ there is is no guild structure to detect and all measures fail at the task, as they should. The ROC and AUC of the equal-time correlation predictor $C_{ij}(0)$ shows that it performs only slightly better than random guessing (AUC = 0.5), for all values of $q$. Taking its absolute value improves prediction beyond random guessing, but is still worse than the spectral measures $\mathcal{P}$ and $\mathcal{TC}$.  

\subsection{Spectral methods work for general ECRM systems with larger sizes}

In natural ecological systems, both the number of guilds and the number of species in each guild can be large. How well can the dynamical pairwise observables predict the resource utilization overlap for different system sizes?  We now consider general ECRMs with larger number of guilds. To explore finite size scaling, we consider three families of ECRMs with numbers of species, resources, and guilds given in Table~\ref{tab:fss}. 

As we increase the size of the ECRMs by increasing the species size $N$, we also increase the number of resources in proportion, $M\propto N$, and the number of guilds as $k\propto \sqrt{N}$.
We take $p_\text{in} = \left(1/2+q\right)\left(N/6 \right)^{-1/2}$ and $p_\text{out} = \left(1/2-q\right)\left(N/6 \right)^{-1}$.
The scalings with $N$ are chosen such that the average connectivity degree for each species in the effective resource-utilization matrix $\cct$ becomes constant at large $N$ (see \textbf{Methods}~\ref{sec:si:scaling} for details). The particular prefactors are chosen so that the numbers are consistent with the $N=6$ case considered earlier.

\begin{table}
\begin{tabular}{|c|c|c|}
\hline
$N$ &  $M \propto N$&$k \propto \sqrt{N}$\\
\hline
 6 & 30 & 2 \\ \hline 
 24 & 120 & 4 \\ \hline 
 96 & 480 & 8 \\ \hline
\end{tabular}
\caption{Parameters for the ECRM with different system size.}
\label{tab:fss}
\end{table}

\begin{figure*}
\includegraphics[width=1\linewidth]{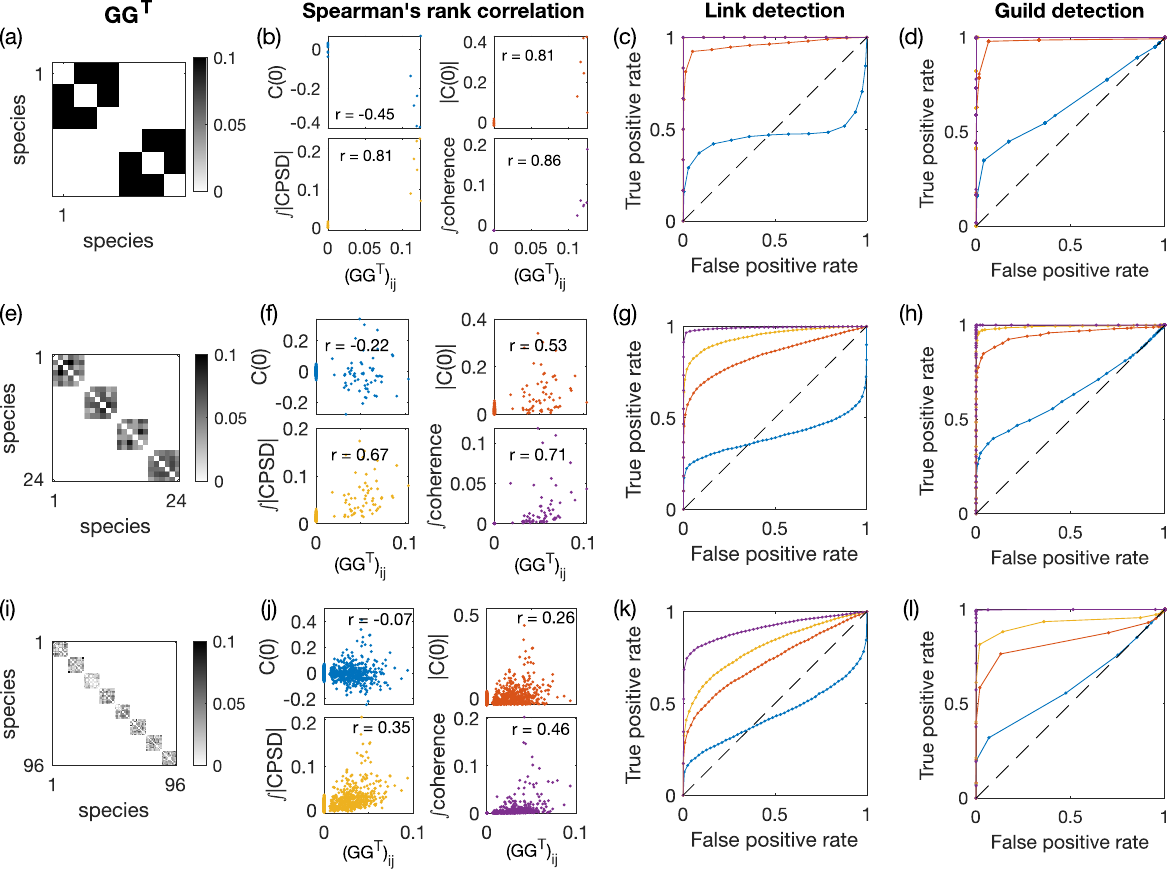}
\caption{\textbf{Pairwise observables for the simulated ECRM ecosystem of different sizes} (Table~\ref{tab:fss}), with $q=0.5$. 
(a) Effective resource-utilization matrix $\cct$ with $N = 6$ species, $M = 30$ resources, $k = 2$ guilds, and $q=0.5$. (b) Correlation between pairwise measures of the species abundances and the resource utilization overlap. Measured pairwise observables include equal-time correlation $\cz$ and its absolute value $\lvert  \cz \rvert$, the total absolute value of CPSD, and the total coherence. (c) ROC curve of link detection for the four pairwise observables. Same color code as panel (b); note that the total absolute value of CPSD overlaps with the total coherence, as both measures return perfect predictions. (d) Same as (c) but for guild detection, after performing single linkage clustering on the pairwise metrics with a tunable threshold. (e-h) Same as (a-d) but for $N = 24$, $M = 120$, and $k = 4$. (i-l) Same as (a-d) but for $N = 96$, $M = 480$, and $k = 8$.
For all ECRMs, the environment is driven by OU processes with the intrinsic timescale set to $\omega_\alpha = 1$. For all system sizes, the total duration of the simulated trajectory is set to $t_f = 20000$, and the sampling timestep is set to $\Delta t_\text{sampling} = 1$.
}
    \label{fig:ecrm_fss_qp5}
\end{figure*}

We monitor the predictive performance of the same four  pairwise measures as before ($\cz$, $\lvert \cz \rvert$, $\mathcal{P}$, and $\mathcal{TC}$) as we increase the system size from $N = 6, k = 2$ to $N = 24, k = 4$, and then to $N = 96, k = 8$ (Fig.~\ref{fig:ecrm_fss_qp5}(a,e,i)). 
We compute Spearman's correlation coefficient between the pairwise observables and the elements of the $\cct$ matrix (Fig.~\ref{fig:ecrm_fss_qp5}(b, f, j)),  
and the ROC of guild detection after single linkage clustering (Fig.~\ref{fig:ecrm_fss_qp5}(d,h,i)).

By the choice of scaling for $p_{\rm in}$ and $p_{\rm out}$, the $\cct$ matrix is sparse for all system sizes.
We can set a threshold on our measures to perform link detection, i.e. identify whether there exists an effective interaction between species due to competition for the same resource.
As shown in Fig.~\ref{fig:ecrm_fss_qp5} for $q=0.5$ (and SI Fig.~\ref{fig:ecrm_fss_qp4} for $q=0.4$), for all the three system sizes tested, the total coherence remains the best among the four predictors. 
The summary statistics in Fig.~\ref{fig:ecrm_fss_stat} show that the total coherence always performs the best among the four predictors, although all get worse as the system size increases. {\RV{For $q = 0.5$,}} guild detection (Fig.~\ref{fig:ecrm_fss_stat}(c)) has a better AUC than link detection (Fig.~\ref{fig:ecrm_fss_stat}(b)), especially for large system sizes, thanks to error correction afforded by single-linkage clustering. {\RV{For $q = 0.4$, however, guild detection performance is worse at larger system size, as single-linkage clustering over-corrects given the presence of non-zero resource competition across different guilds (Fig. S1(d,i)). Alternatively, one can perform single-linkage clustering but only for one iteration, such that immediately adjacent neighbors are now linked. While suggesting more sophisticated community-detection methods, this middle ground provides a better recovery of the guild structure compared to link detection and guild detection with full single-linkage clustering (Fig. S1(e,j)).  }}

\subsection{Reconstruction of the $\cct$ matrix depends on sampling rate and total time series length}

The success of CPSD and coherence in detecting guild structures depends on the interplay of multiple timescales. These timescales include the doubling timescale $(\gamma r)^{-1}$ of the species, their typical lifetime $d_x^{-1}$,
the environmental fluctuation timescales $\omega_\alpha^{-1}$, and their relaxation time $d_R^{-1}$. Additionally, there are experimental timescales: the total time $t_f$ of data acquisition and sampling interval $\Delta t_{\rm sampling}$.
Here, we discuss the effects of the finite time series and sampling rate on recovering the elements and structure of the $\cct$ matrix. 

\begin{figure}
\includegraphics[width=0.45\textwidth]{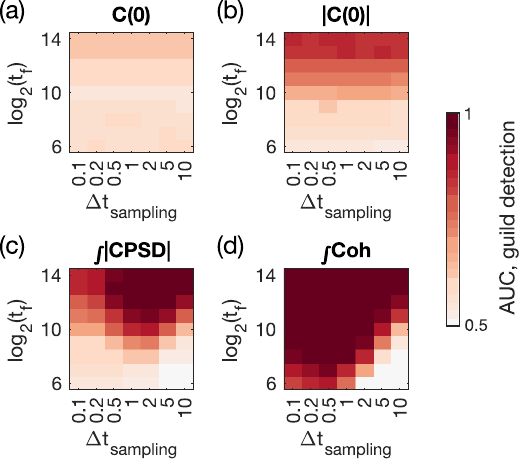}
\caption{\textbf{Influence of finite data duration and sampling rate on guild detection using pairwise measures of species abundance.} AUC of the guild detection task using (a) Equal-time correlation $\cz$,  (b) its absolute value $\lvert  \cz \rvert$, (c) the total absolute value of CPSD, and (d) the total coherence, as the function of sampling duration $t_f$ and interval $\Delta t_{\rm sampling}$. For reference the species turnover timescale is $d_x^{-1}=3.33$. 
The results are averaged over 50 random realizations of the ECRMs with $N = 6$, with OU environmental drive at $\omega_\alpha = 1$ and guild-structure bias $q = 0.4$. }
\label{fig:tf_tsampling}
\end{figure}

 As we increase the total time of acquisition $t_f$, the accuracy of guild detection  improves for all 4 observables as measured by AUC, as expected (Fig.~\ref{fig:tf_tsampling}(a)-(d)).
For equal-time measurements ($\cz$ and $\lvert \cz \rvert$, (Fig.~\ref{fig:tf_tsampling}(a), (b)), the sampling interval $\Delta t_{\rm sampling}$ does not change the prediction significantly. However, for the total absolute CPSD and the total coherence (Fig.~\ref{fig:tf_tsampling}(c), (d)),  the AUC for guild detection shows an optimal near $\Delta t_\text{sampling} = 2$ and $\Delta t_\text{sampling} = 0.5$ respectively, for which the AUC for guild detection is largest for the biggest range of total data acquisition timescale.
For comparison, in our simulation the species death rate is $d_x = 0.3$, corresponding to a turnover timescale of 3. In the limit of small $t_f$ and large $\Delta t_\text{sampling}$, the number of time points is less than the number of the time points required by spectral method given the  frequency resolution we specified (for FFT the number of data points needs to be at least double the number of the frequency resolution), which means the methods cannot work (bottom right corner of Fig.~\ref{fig:tf_tsampling}(d)). Similar results hold for Spearman's correlation between the pairwise metrics and the $\cct$ matrix element (SI Fig.~\ref{fig:tf_tsampling_spearman}).

In summary, predictions always benefit from longer time traces, but there exists an optimal, finite sampling frequency for the two spectral methods, which perform best.

\subsection{Reconstructing guild structure from relative species abundances}
When dealing with genomic data, often only  relative abundances of the species rather than absolute abundances are available. We tested whether the CPSD and coherence observables remain useful for species structure inference from relative abundances, for systems of different sizes. 

SI Fig.~\ref{fig:relative_abundance} shows that, for small number of species ($N=6$ and $24$), total coherence has poor predictive power when using relative abundances, while the total magnitude of CPSD still performs well. This result is surprising since coherence as a function of frequency has more structure than CPSD. However, total coherence outperforms other measures for large numbers of species ($N = 96$).
This result is consistent with previous work~\cite{friedman_inferring_2012} where the impact of relative abundance was reported to be less severe when the number of species is large.

{\RV{
\subsection{Reconstruction of guild structure with heterogeneity in guild size and evenness}

So far we have examined ECRMs where guild size is uniform across a community. Here, we test whether the dynamical observables still perform well when there is heterogeneity in guild size or unevenness in species abundances. 

SI Fig.~\ref{fig:si:hetero_guild} shows that the ability to successfully detect links remains high for communities comprised of guilds of variable size. Coherence remains the best predictor among the four pairwise measures. Guild detection, however, becomes worse when the ECRM allows for the sharing of resources across different guilds ($q=0.4$). Detectability depends on guild size: while link detectability is unaffected by guild structure, smaller guilds are harder to detect after single linkage clustering.

Another source of heterogeneity is species evenness, i.e. certain species can be more abundant in the community. To allow variability in species abundances, especially across different guilds, we adjust the number of resources associated with each guild. 
SI Fig.~\ref{fig:si:hetero_evenness} shows that the local guild detection is improved by species unevenness, especially if species in small guilds are present at high relative abundance. While local detectability isn't monotonic with species evenness, higher relative abundance improves AUC. Thus, a small guild (e.g., two species) in a community with larger guilds may be misidentified—unless its members are highly abundant.}}

\begin{figure*} 
\includegraphics[width=0.9\textwidth]{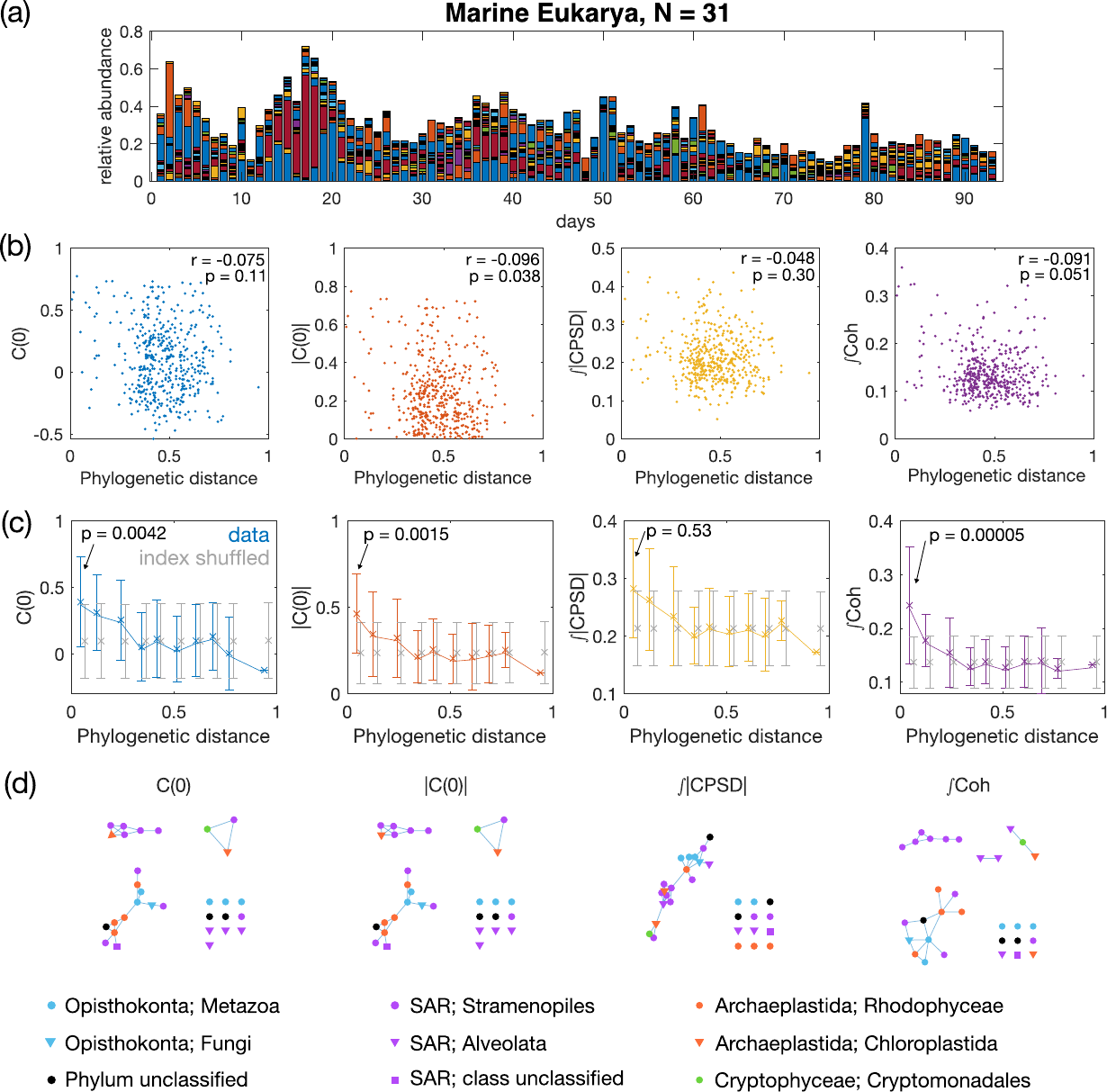}
\caption{{\bf Pairwise metrics of the relative abundances of marine eukaryotic OTUs correlate with phylogeny.}  (a) {\RV{Relative abundance of the marine eukaryotic OTUs (downsampled to $N = 31$, see text) over 93 days. Each color represents one of the 31 OTUs. Data are from~\cite{marine_data}.}}  (b) Scatterplot of pairwise metrics of relative abundance exhibit a weak anti-correlation with the pairwise sequence distance (Jukes-Cantor distance). Pairwise measures include the equal-time correlation $\cz$ and its absolute value $\lvert \cz \rvert$, the total magnitude of CPSD, and total coherence. Spearman's rank correlation coefficient $r$ and the $p$-value are given in each panel. (c) 
Binned average of the pairwise metrics vs. pairwise sequence distances. Binning is taken at regular intervals of the sequence distance.
The mean across each bin is plotted with a cross symbol {\RV{(connected with lines to guide the eye)}}, and error bars given the standard deviation within each bin. 
In grey are shown the results from a randomized control, where the order of indices of eukaryotic OTUs are shuffled randomly (number of shuffling = $10^5$). The \textit{p}-value for the mean of the pairwise metrics in the first bin to be different from the index shuffled null model is given in each panel (see \textbf{Methods}~\ref{methods:pval} for details). 
(d) Resource overlap networks inferred using the four pairwise metrics, with threshold set to produce $k = 13$ clusters including singletons. The phylum and class of each OTU is given by respectively the color and the shape of the node, and in the legend with the format \textit{phylum};\textit{class}.} 
\label{fig:marine}
\end{figure*}

\subsection{Applying temporal metrics to marine data}

To illustrate the usefulness of CPSD and coherence on real data, we tested our 4 pairwise metrics on a public dataset consisting of time-series of relative abundance of a total of 49,637 bacterial and eukaryotic operational taxonomic units (OTUs) from coastal plankton collected over 93 consecutive days~\cite{marine_data}. For each of the 93 days we have 3 samples for each OTU, and we use the averaged relative abundance across the 3 samples as the relative abundance for each OTU. As the spectral methods have so far been developed considering consecutive time points sampled at regular intervals, we choose to analyze only the eukaryotic OTUs, because the bacterial data contained missing days across all OTUs. Furthermore, since our methods have been developed and tested on synthetic data without zero abundances, we limit our analysis to the $N = 31$ eukaryotic OTUs which are present on each of the 93 days. As the total number of eukaryotic OTUs is much larger than the considered subset, we expect the total coherence and the total magnitude of CPSD should  not be adversely effected by the use of relative abundances.

We use the time series of the relative abundances of the $N=31$ eukaryotic OTUs to compute the four pairwise metrics as described in the previous sections: the equal-time correlation $\cz$  its absolute value $\lvert\cz\rvert$, the total magnitude of CPSD $\mathcal{P}$, and the total coherence $\mathcal{TC}$. 
Ideally, we would like to compare the pairwise metrics with the elements of the resource utilization overlap matrix $\cct$. However, since there is no \textit{a priori} measurement or knowledge of the guild structure in natural communities, we use phylogenetic distance as a proxy, following previous work~\cite{sireci_environmental_2023}. The intuition is that the closer the species are genetically, the more likely they are to share the resources~\cite{martiny_phylogenetic_2013}. 

Comparing the pairwise metrics of the species abundances and the pairwise phylogenetic distance (Fig.~\ref{fig:marine}(a)),  all four pairwise metrics exhibit a weak negative correlation with the phylogenetic distance, indicating that more closely related taxa are more likely to share resource utilization capabilities, which is consistent with the literature~\cite{sireci_environmental_2023, lemos_phylogeny_2024}. After binning the phylogenetic distances using regular intervals and computing the mean and average of the pairwise metrics in each bin, the trend in negative correlations becomes more visible (Fig.~\ref{fig:marine}(b), \textit{colored bars}). To examine whether this negative correlation is statistically significant, we created a null model by destroying any phylogenetic correlation by randomly shuffling the indices of the {\RV{OTUs}} before computing the pairwise metrics. The randomization is performed $10^5$ times to collect statistics. As shown by the gray error bars in Fig.~\ref{fig:marine}(b), the null model results in flat pairwise metrics for all metrics, as a function of the phylogenetic distance. At large phylogenetic distances, the data and the index-shuffled null model return the same mean and variance, as indicated by the overlaying errorbars in Fig.~\ref{fig:marine}(b). 

Among the four metrics, total coherence exhibits the biggest difference between the data and the null model for small phylogenetic distances, with a $p$-value of $5\times 10^{-5}$ for the test of the mean in the first bin against the index-shuffled null model (see \textbf{Methods}~\ref{methods:pval} for details).
These results indicate that even for data collected experimentally in nature, total coherence works well in distinguishing structures in OTUs sharing common resources.  

With the pairwise metrics, now we can infer guild structures among the $N = 31$ eukaryotic OTUs using single-linkage clustering with a tunable threshold on the metrics(SI Fig.~\ref{fig:si:euk31_matrix}(a)). As there is no \textit{a priori} knowledge of how those eukaryotic OTUs share resources, i.e. no ground truth to compare the guild structure to, we take the strategy of comparing the inferred structures across the four pairwise metrics. 
The threshold for each metric is selected such that the resulting number of clusters, including singletons, is the same for all four pairwise metrics ($k = 13$). The thresholded connections can be found in SI Fig.~\ref{fig:si:euk31_matrix}(b).

As shown by Fig.~\ref{fig:marine}(c), different metrics yield different predicted guild structures.
Clustering based on coherence gives two non-trivial clusters that exclusively contain OTUs of the same class, which does not occur for other metrics. The equal-time correlation $\cz$ and its absolute value $\lvert \cz \rvert$ return similar networks. Meanwhile, the total absolute CPSD gives a different network with one giant component and many singletons. This further suggests that the coherence metrics provide a useful predictor of species functional relationships.

\section{Discussion}

We have shown that by judicious analysis of abundance dynamics data, one can reliably learn patterns of resource competition in microbial communities. Our success emerged from three key perspectives that distinguish this study from prior work.

First, rather than attempting to infer effective interactions between taxa such as those described by a Lotka-Volterra framework~\cite{stein_ecological_2013}, we focus on resource-mediated interactions. Consumer resource models have been shown to quantitatively predict abundance and resource dynamics in communities~\cite{gowda_genomic_2022,ho_resource_2024}. Further, resource-explicit formalisms do not suffer from the ambiguities of effective interactions, especially in the context of microbial communities~\cite{momeni_lotka-volterra_2017}. Moreover, the resource-centric picture of communities naturally motivates us to focus on coarse patterns of resource utilization overlap, which are empirically supported by the existence of guilds in communities. Guilds allow us to simplify the problem from trying to infer every interaction in a community to inferring a lower-dimensional structure that aggregates strains together by resource preferences.

Within this dynamic resource-centric picture, our second key innovation was to move beyond simple correlations ($\cz$) by employing spectral methods that utilize full temporal information. Our method stands in contrast to covariance-based approaches~\cite{spieceasi,friedman_inferring_2012} that estimate covariance or inverse covariance matrices to infer associations but do not utilize dynamics. One exception is a recent study that employs temporal delays and Granger causality to infer associations~\cite{mainali_detecting_2019}. Similarly, spectral metrics (CPSD and coherence) provide a high-fidelity picture of resource overlaps in communities by extracting information from multiple temporal delays. We expect these methods will prove powerful for inferring structure in complex community where the acquisition of high-resolution time series has become commonplace.

Finally, our spectral methods not only embrace but rely on the inherently dynamic nature of the environment ($R_\alpha(t)$). In contrast, time-series methods, including those based on Lotka-Volterra (LV) models~\cite{fisher_identifying_2014}, often assume steady-state dynamics~\cite{ives_estimating_2003,maynard_predicting_2020} (with notable exceptions~\cite{marsland_community_2020,cui_effect_2020}). However, it is empirically clear that microbes live in environments that are inherently dynamic~\cite{birch_effect_1958}. One of the most striking pieces of evidence of the importance of resource dynamics in the microbial world is recent physiological work showing that bacteria dynamically allocate cellular resources in a manner that optimizes growth under changing environments but not steady state conditions~\cite{chure_optimal_2023,balakrishnan_conditionally_2023}. Therefore, we regard the fact that our method relies on resource-driven abundance dynamics as an empirically motivated strength.

Using this approach, our analysis of eukaryotic microbial dynamics in a high-resolution marine dataset showed promising results in two domains. First, we observed that spectral methods outperform simple correlations in detecting statistically significant resource overlap between phylogenetically related taxa (Fig.~\ref{fig:marine}(b)). Second, it is enticing that these methods yield distinct network architectures from the same data (Fig.~\ref{fig:marine}(c)). An important avenue for future work is to more carefully vet the predicted associations from these metrics with synthetic communities.

Despite these successes, there are important avenues for improvement. First, as discussed above, at present, our method does not naturally deal with missing time points, uneven temporal sampling, or zeros in the data. It is an important avenue for future work to merge these methods with recent principled advances for handling zeros in amplicon sequencing data~\cite{callahan_exact_2017} or utilizing Lomb-Scargle periodogram methods for unevenly sampled data~\cite{scargle_studies_1982}.

Finally, our work complements recent efforts to infer consumer-resource models or metabolic parameters directly from sequencing and metabolomic data~\cite{goyal_ecology-guided_2021,li_statistical_2023,gowda_genomic_2022}. The method of Goyal {\em et al.}~\cite{goyal_ecology-guided_2021} attempts to infer a much more detailed resource-exchange network by using genomic and metabolomic data to pinpoint actual cross-feeding interactions. An exciting avenue for future work would be to combine the sophisticated spectral methods developed here with more explicit genomic or metabolomic datasets.

\section{Methods}

\subsection{Initialize parameters of ECRMs}\label{methods:params}

For all system sizes, the parameters of the ECRM systems are set as follows. The uptake rate for non-zero resource intake is set to $r = 0.1$. The yield $\gamma_{i\alpha}$ is given, drawn from a Gaussian distribution with mean 1 and standard deviation 1/6. Redraws are performed to ensure all yields are non-negative. The death rate of the species is set to $d_x = 0.3$, while the depletion rate of the resource is set to $d_R = 0.5$. For sinusoidal drives, the timescales of the environmental fluctuation, $\omega_\alpha$, are chosen randomly from a log-uniform distribution with a pre-determined range to represent the mixture of multiple timescales in real environment and to avoid resonance. The phase $\phi_\alpha$ is chosen from a uniform random distribution between 0 and $2\pi$. For OU environmental drives, thanks to its stochastic nature, we can choose the intrinsic frequencies $\omega_\alpha$ to be the same for all resources $\alpha$.

\subsection{Finite size scaling in ECRMs}\label{sec:si:scaling}

In order to bridge the toy model with a small number of species and guilds with realistic ecological systems with large numbers of species and guilds, we scaled the parameters of the ECRM given different system sizes. As we increase the number of species $N$, we want to also increase both the number of  guilds $k$ and the number of species within each guild $N/k$. Hence, we set the number of guilds $k \propto \sqrt{N}$. The number of resources should also increase as $M \propto N$. Setting the probability of species $i$ to have a non-zero intake of its preferred resource as $p_\text{in}$, and the probability of species $i$ to have a non-zero intake of its non-preferred resource as $p_\text{out}$, we can compute the expectation value of the average degree of each species in the effective species-species interaction due to resource utilization overlap, the expectation value of $(\cct)_{ij}$, etc. 

\textbf{Case 1: species $i$ and $j$ belong to the same guild}.
The probability that there is no link between two species is equal to the probability that there is no common resource between the two, which is 
\beq
P((\cct)_{ij} = 0) = \left(1-p_\text{in}^2\right)^{M/k}\left(1-p_\text{out}^2\right)^{M(1-1/k)}.
\eeq

\textbf{Case 2: species $i$ and $j$ do not belong to the same guild}.
The probability that there is no link is 
\beq
P((\cct)_{ij} = 0) = \left(1-p_\text{in} p_\text{out}\right)^{2M/k}\left(1-p_\text{out}^2\right)^{M(1-2/k)}.
\eeq

Given those two cases, the average degree for each node is
\beq
\begin{split}
d & = \left( 1 - (1-p^2_\text{in})^{M/k} (1-p^2_\text{out})^{M(1-1/k)} \right) \left(\frac{N}{k}-1\right)  \\
& + \left( 
1 - (1 - p_\text{in} p_\text{out})^{2M/k} 
(1 - p^2_\text{out})^{M(1-2/k)} 
\right) 
N
\left(1 - \frac{1}{k}\right). 
\end{split}
\eeq
For proper scaling, we want that, in the limit of large $N$, the average degree $d$ converges to a constant value. The choice $p_\text{in} \propto N^{-1/2}$ and $p_\text{out} \propto N^{-1}$ satisfies this condition, by ensuring that the average degrees towards in-guild and out-of-guild species are both finite. Notice that this scaling results in a $1/N$ scaling for both the expectation value and the variance of elements of the $\cct$ matrix. 

{\RV \subsection{Initiating guild structure with guild-size variability}

To introduce more realistic heterogeneity in the ECRMs, we introduce a tuning parameter $q_N$ which sets geometrically-spaced guild sizes, $N_k = q_N^{k-1} N_1$, and $\sum N_k = N$.  Furthermore, to allow variability in species abundance, especially across different guilds, we introduce another tuning parameter $q_M$, which accordingly adjusts the number of resources associated within each guild, again by setting a geometric series, where $M_k = q_M^{k-1} M_1$, and $\sum M_k + N = M$. $N_k$'s and $M_k$'s are rounded up to the nearest integer. If the sum of the rounded up number is smaller than the total number of species $N$, the missing species is added to the largest guild. If the sum of the rounded up number is greater than the total number of species $N$, the additional species is subtracted from the smallest guild, unless if the smallest guild has only 2 species, then the additional species is subtracted from the second smallest guild, such that the minimum size of each guild is 2.

To give intuition, if $q_N = 1$, all guild sizes are equal. The further $q_N$ is from 1, the greater the differences in guild sizes. If $q_M = q_N$, then on average each species is exposed to the same number of resources, and the species are even. 
}

{\RV \subsection{Local measure of detectability}

We introduce a local measure of detectability. Namely, for guild $k$, the true positive rate is the ratio between the number of detected adjacency links between all species present in guild $k$ and the number of possible pairs, $N_k(N_k-1)/2$. The guild-specific false positive rate is defined as the probability to falsely classify a link between a species $i$ in guild $k$ with a species $j$ outside of guild $k$.}

\subsection{Measure sequence distance in marine eukaryotic OTUs}\label{methods:seqdist}
To measure sequence distance, we first align the 18S eukaryotic rRNA sequences of the $N_\text{sub} = 31$ eukaryotic OTUs using the Multiple alignment program for amino acid or nucleotide sequences (MAFFT)~\cite{mafft}. Then, sequence distance is computed for each pair of the OTUs, using the built-in MATLAB function seqpdist, namely the Jukes-Cantor distance.

\subsection{Statistical tests for binned pairwise metrics against index shuffled null model}\label{methods:pval}

For analyzing the structure in the marine eukaryotic OTU timeseries data, we conduct statistical tests for the binned pairwise metrics against the index shuffled null model. Specifically, we shuffle the index of species randomly and then compute the pairwise measurements. This null model preserves the set of all elements of the pairwise measures, but destroys any association with the species relationships.

The statistical test conducted in Fig.~\ref{fig:marine}(b) checks  whether the mean of the 8 pairwise measurements in the first bin are significantly larger than in the index shuffled randomization. We performed index shuffling $10^5$ times. For each randomization, we computed the mean in the first bin. 
We computed the $p$-value as the proportion of shuffling experiments that yielded a larger mean in the first bin than the observed one.

\section*{Code availability}
Codes and data generating all figures in this manuscript are available on OSF (https://doi.org/10.17605/osf.io/hvnjt).

\section*{Acknowledgments}
This work was partially supported by the European Research Council Consolidator Grant n. 724208. This work was also supported by the Bettencourt Schueller Foundation to T.M. X.C. acknowledges support by the L'Or\'{e}al-UNESCO Young Talents France Fellowship. S.K. acknowledges the National Institute of General Medical Sciences R01GM151538. S.K. acknowledges a CAREER award from the National Science Foundation (BIO/MCB 2340416) and support from the National Science Foundation through the Center for Living Systems (grant no. 2317138).  S.K. acknowledges financial support from the National Institute for Mathematics and Theory in Biology (Simons Foundation award MP-TMPS-00005320 and National Science Foundation award DMS-2235451).

\bibliographystyle{pnas}

 % si_revision
\setcounter{figure}{0}
\makeatletter 
\renewcommand{\thefigure}{S\@arabic\c@figure}
\makeatother

\onecolumngrid
\newpage

\section*{Supplementary Information}

\begin{figure}[h]
    \includegraphics[width=1\linewidth]{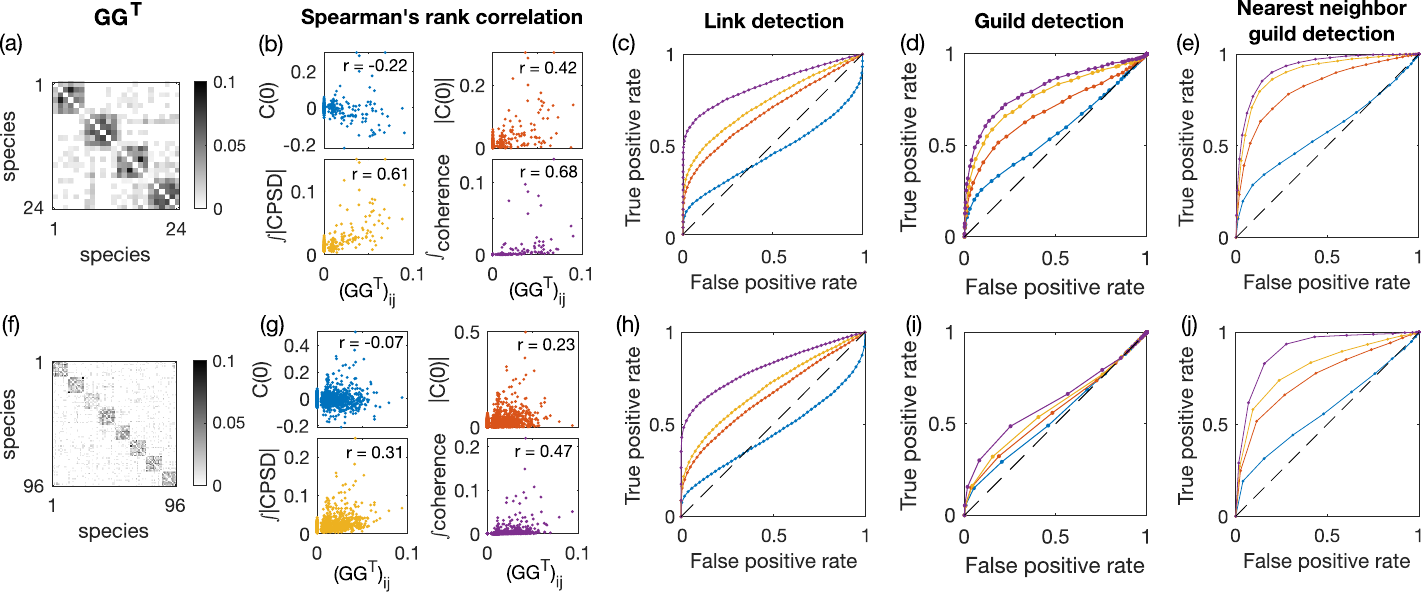}
 \caption{\textbf{Pairwise observable analysis for the simulated ECRM ecosystem of different sizes.} Same as Fig.~\ref{fig:ecrm_fss_qp5} but with $q=0.4$.}
    \label{fig:ecrm_fss_qp4}
\end{figure}

\begin{figure*}[h]
\includegraphics[width=0.7\textwidth]{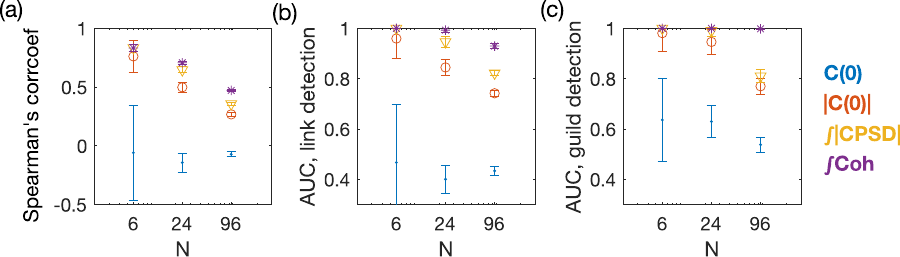}
\caption{\textbf{Coherence and CPSD outperforms equal-time correlation measures in recovering the resource-utilization overlap matrix $\cct$ for all sizes of systems.} 
(a) Spearman's rank correlation coefficient between the four considered pairwise observables (equal time correlation $\cz$ and its absolute value $\lvert \cz \rvert$, the total magnitude of CPSD $\mathcal{P}$, and the total coherence $\mathcal{TC}$) and the elements of the $\cct$ matrix, for three families of ECRM systems with $N = 6, 24, 96$ defined in Table~\ref{tab:fss}. Error bars represent standard deviation across 100 random realizations of species abundances  for systems with $N = 6$ and with $N = 24$ (10 random ECRM systems each with 10 random OU environmental drives with $\omega_\alpha = 1$), and across 10 random realizations of species abundances  for systems with $N = 96$ (10 random ECRM systems each with 1 random OU environmental drive with $\omega_\alpha = 1$). 
(b) Area under curve (AUC) for link detection for the same systems as in (a). Each AUC is computed for the same ECRM system, averaged over the ROC of all 10 random OU drives. 
(c) AUC for guild detection after single linkage clustering for the same systems as in (a). Guild-structure bias is $q = 0.5$.
}  
\label{fig:ecrm_fss_stat}
\end{figure*}

\begin{figure}[h]
\includegraphics[width=0.45\textwidth]{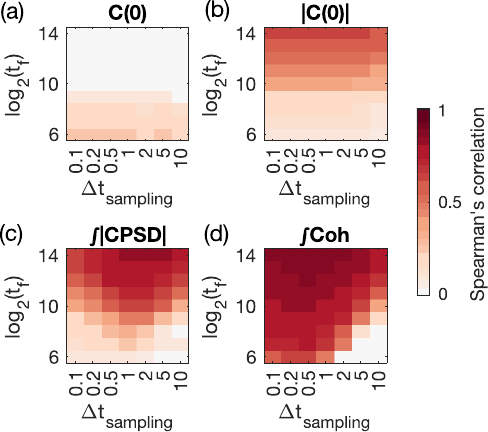}
\caption{
\textbf{Influence of finite time-series duration and sampling rate on Spearman's correlation coefficient between the pairwise measures of species abundance and the resource utilization overlap matrix $\cct$.} Spearman's correlation coefficient (colorbar) between elements of the $\cct$ matrix and elements of (a) Equal-time correlation $\cz$,  (b) its absolute value $\lvert \cz \rvert$, (c) the total absolute value of CPSD, and (d) the total coherence, as the function of sampling duration $t_f$ and the sampling interval $\Delta t_{\rm sampling}$. The results are averaged over 50 random realizations of the ECRMs with $N = 6$, with OU environmental drive with $\omega_\alpha = 1$ and guild-structure bias $q = 0.4$.}
\label{fig:tf_tsampling_spearman}
\end{figure}

\begin{figure}[h]
\includegraphics[width=0.7\textwidth]{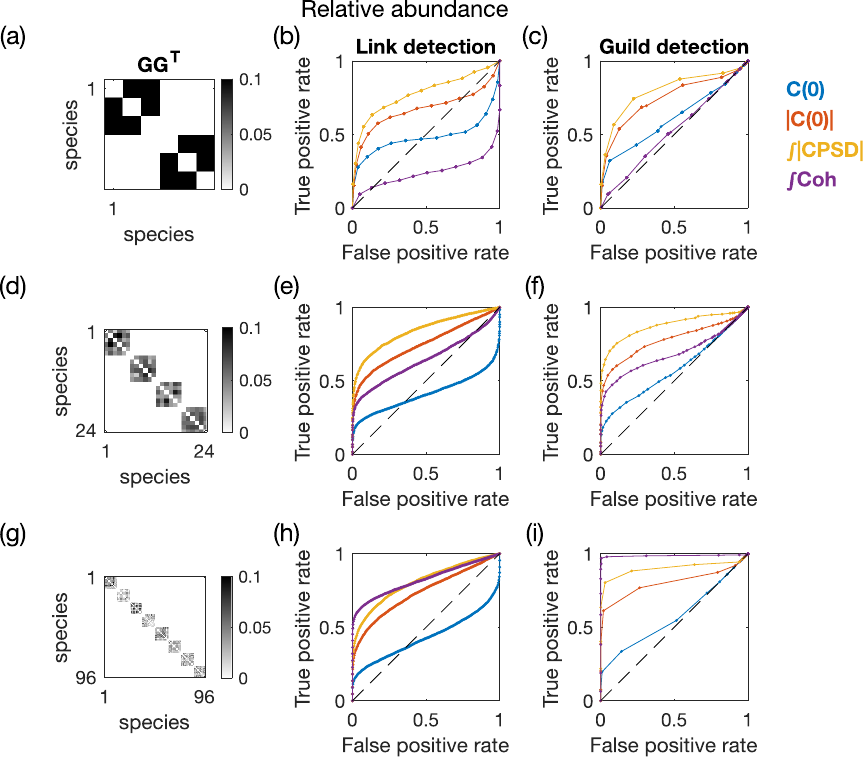}
\caption{
\textbf{Link detection and guild detection of the resource utilization overlap matrix $\cct$ using pairwise measurements of the relative abundance (instead of absolute abundance) of species concentration, for simulated ECRM ecosystems of different sizes}
(Table~\ref{tab:fss}), with guild-structure bias $q=0.5$. 
(a) Effective resource-utilization matrix $\cct$ with $N = 6$ species, $M = 30$ resources, $k = 2$ guilds, and $q=0.5$. (b) ROC curve of link detection for the four pairwise observables. Legend for the colors is given next to panel (c). (d) Same as (c) but for guild detection, after performing single linkage clustering on the pairwise metrics with a tunable threshold. (e-h) Same as (a-d) but for $N = 24$, $M = 120$, and $k = 4$. (i-l) Same as (a-d) but for $N = 96$, $M = 480$, and $k = 8$.
For all ECRMs, the environment is driven by OU processes with the intrinsic timescale set to $\omega_\alpha = 1$. For all system sizes, the total duration of the simulated trajectory is set to $t_f = 20000$, and the sampling timestep is set to $\Delta t_\text{sampling} = 1$. }
\label{fig:relative_abundance}
\end{figure}

\begin{figure}[h]
    \centering    \includegraphics[width=1\linewidth]{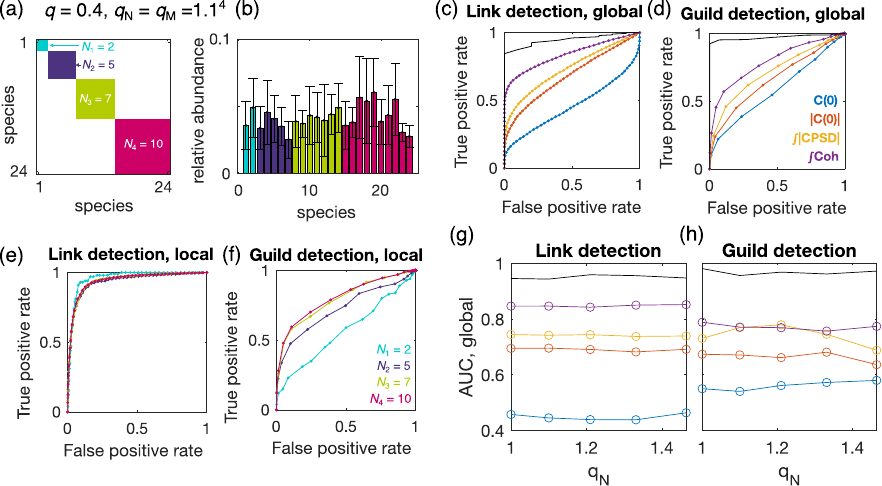}
    \caption{{\RV{\textbf{Link detection and guild detection for heterogeneous ECRM ecosystems with different guild sizes.}
    The guild sizes follow a geometrical series with the common ratio set by the guild-size variability $q_N$ (see \textbf{Methods}). The average number of resources per species is tuned by $q_S = q_M/q_N$, where $q_M$ is the common ratio of the geometrical series of the number of resources associated with each guild. For this plot, $q_S = 1$, which leads to each species being equally presented. 
    Guild-structure bias is set to $q = 0.4$. (a) Schematics for an ECRM with $N = 24$ species, $M = 120$ resources, and $k = 4$ guilds. The guild structures at $q_N = 1.1^4$. The number of species in each guild is $N_1 = 2$, $N_2 = 5$, $N_3 = 7$, $N_4 = 10$, respectively.
    (b) Averaged relative abundance for each species. Color represents distinct guilds. Error bars represent standard deviation across 10 random realizations of the $\cct$ matrix, each with 10 random OU environmental drives with $\omega_\alpha = 1$. (c-d) ROC curve of link detection (c) and guild detection (d) for the four pairwise observables. Legend for the colors is given in panel (d). The black curve is the ROC curve if one uses the ground truth $\cct$ matrix as the binary classifier. Coherence remains the best predictor of the link and guild structure.
    (e-f) Guild-specific local detectability for link detection (e) and guild detection after single linkage clustering (f), using coherence as the binary classifier. Local detectability measures for each guild, the probability of two species correctly being identified in the guild (see detailed definition in \textbf{Methods}). Colors correspond to the four guilds with different sizes as given in panel (a). Although link detection does not depend on guild structure, guild detectability after single linkage clustering shows a considerable difference between guilds of different sizes. In particular, the smallest guild with $N_1 = 2$ becomes difficult to detect.  
    (g-h) Link detection (g) and guild detection (h) for ECRM ecosystems with a range of guild-size variability. 
    Here, we specify $q_N = 1, 1.1^{1}, \dots, 1.1^4$. $q_S = 1$ always, such that the relative abundance for each species is even. Guild sizes are equal at $q_N = 1$, which corresponds to the discussions in the main text. }}}
    \label{fig:si:hetero_guild}
\end{figure}

\begin{figure}
    \centering
    \includegraphics[width=0.65\linewidth]{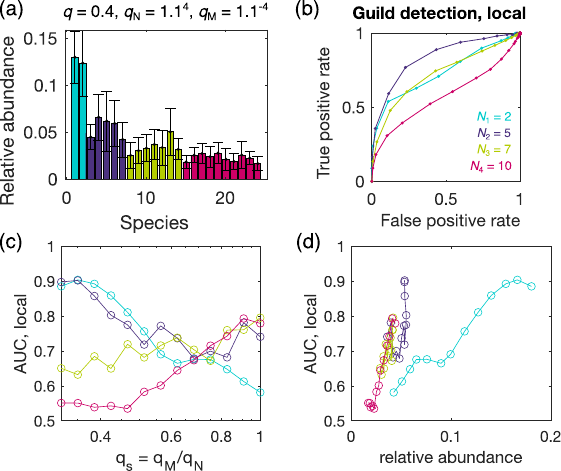}
    \caption{{\RV \textbf{Guild detection for heterogeneous ECRM ecosystems with different guild sizes and different species abundance. Detection of small guilds is rescued by large relative abundance in those guilds.} Shown here are the results for an ECRM ecosystem with $N = 24$ species, $M = 120$ resources, $k = 4$ guilds, and guild-size variability $q_N = 1.1^4$ as exemplified by the schematics in Fig.~\ref{fig:si:hetero_guild}(a). Averaged number of resources per species is tuned by $q_S = q_M/q_N$.  
    (a) Relative abundance of the species in an example ECRM ecosystem, with $q_M = 1.1^{-4}$, i.e. $q_S = q_M/q_N = 1.1^{-8}$. This community is uneven. Species belonging to the smallest guild with size $N_1 = 1$ have the largest relative abundance.  Colors represent distinct guilds. Error bars represent standard deviation across 10 random realization of the $\cct$ matrix, each with 10 random OU environmental drives with $\omega_\alpha = 1$. (b) Guild-specific local detectability for guild detection, using coherence as the binary classifier. Curves with different colors correspond to the four different guilds. The detectability of the smallest guild with size $N_1 = 2$ (cyan) improves compared to the ECRM with the same guild sizes but even species abundance (Fig.~\ref{fig:si:hetero_guild})(f). (c)  
     AUC for local guild detectability using coherence as the binary classifier (Area under the curves in panel (b)), for various species unevenness with the tuning parameter $q_S = 1.1^{-12}, 1.1^{-11}, \dots, 1.1^{-1}, 1$. The value $q_S = 1$ corresponds to results presented in Fig.~\ref{fig:si:hetero_guild}, where species are even. (d) AUC for local guild detectability against the mean relative abundance of all species in specific guilds. As the relative abundance increases, the AUC also increases, suggesting that guilds are more detectable if species in the guild are more abundant. Specifically, at small values of $q_S$, i.e when the smaller guilds have large relative abundance and the large guilds have small relative abundance, it becomes more difficult to detect the larger guilds ($N_4 = 10$ curve in panel (c)), although the AUC is still greater than 0.5.  
     Results in (c) and (d) are averages over 10 random realizations of the ECRM systems.  }
    }
    \label{fig:si:hetero_evenness}
\end{figure}

\begin{figure}[h]
    \centering   \includegraphics[width=1\linewidth]{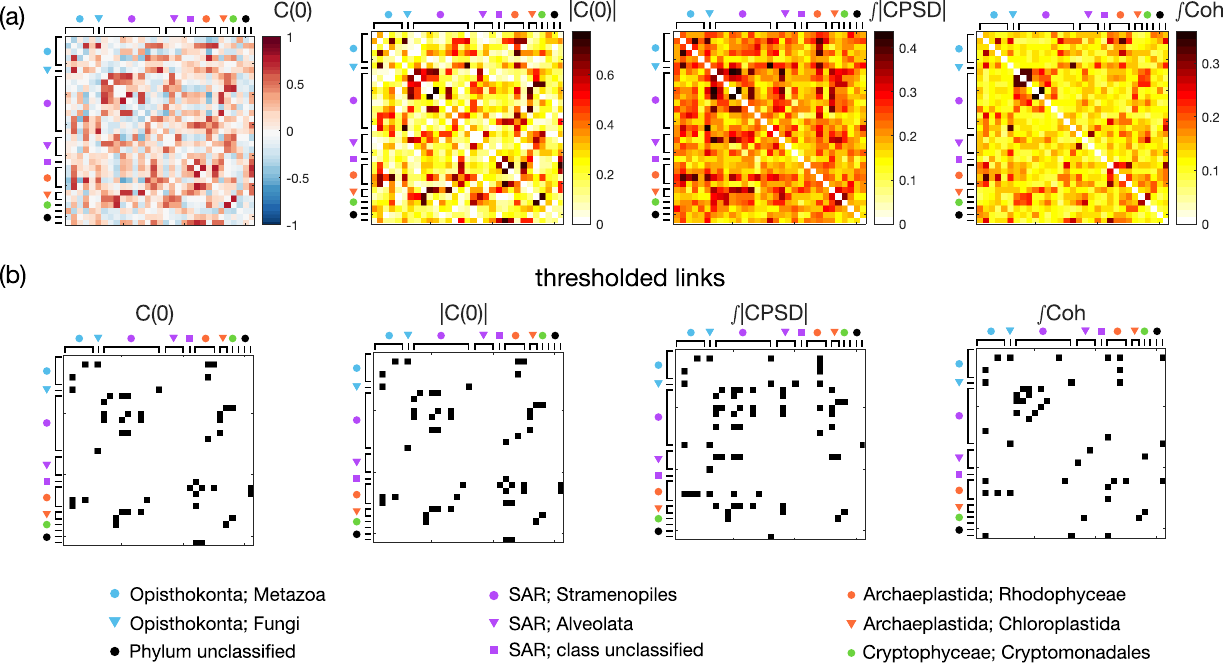}
    \caption{\textbf{Pairwise measurements of the 93-day time series of relative abundance of eukaryotic OTUs in the Martin-Platero et al. coastal plankton dataset} (Fig.~\ref{fig:marine}, data from~\cite{marine_data}, downsampled to $N = 31$). (a) Equal-time correlation $\cz$ and its absolute value $\lvert \cz \rvert$, the total magnitude of CPSD, and the total coherence of the time-series of relative abundance of marine eukaryotic OTUs ($N = 31$). The phylum and the class of each OTU is indicated by an unique color and shape (given in legend at bottom). (b) The adjacency matrix formed by thresholding the pairwise measurements in panel \textit{a} to form $k = 13$ clusters (as in Fig.~\ref{fig:marine}(c) and described in text). Black indicates a link, while white indicates no link. }
    \label{fig:si:euk31_matrix}
\end{figure}

\end{document}